# Gamma-Ray Burst Peak Duration as a Function of Energy

E. E. Fenimore, and J. J. M. in 't Zand

Los Alamos National Laboratory, MS D436, Los Alamos, NM 87545

E-mail: efenimore@lanl.gov

and

J. P. Norris, J. T. Bonnell, and R. J. Nemiroff

Goddard Space Flight Center, Greenbelt, MD 20771

## ABSTRACT

Gamma-ray burst time histories often consist of many peaks. These peaks tend to be narrower at higher energy. If gamma-ray bursts are cosmological, the energy dependence of gamma-ray burst time scales must be understood in order to correct the time scale dependence due to the expansion of the universe. By using the average autocorrelation function and the average pulse width, we show that the narrowing with energy follows, quite well, a power law. The power law index is $\sim$ -0.4. This is the first quantitative relationship between temporal and spectral structure in gamma-ray bursts. It is unclear what physics causes this relationship. The average autocorrelation has a universal shape such that one energy range scales linearly with time into all other energy ranges. This shape is approximately the sum of two exponential.

*Subject Headings:* Gamma-Rays: Bursts

# 1. INTRODUCTION

The Burst and Transient Experiment (BATSE) on the Compton Gamma Ray Observatory has deepened the mystery of gamma-ray bursts (GRBs) rather than solving it. GRBs appear to be isotropic on the sky yet there is a dearth of faint events compared to the brightest events (Meegan et al. 1992). The two most likely explanations for this situation are the bursts are at cosmological distances (and the dearth of events is due to effects associated with the expansion of the universe) or the events are from an extended halo about our Galaxy (and the dearth of events is due to a decrease in the density of neutron stars in the halo). If cosmological, the expansion of the universe shifts the photon energies by a factor of $1/(1 + z)$ where $z$ is the redshift and stretches the temporal structure by a factor of $1 + z$. Indeed, time dilation has been claimed on all time scales within GRBs (Norris et al. 1994, 1995a, Davis et al. 1994). A variety of tests have been used to detect the time dilation. The total-count test, wavelet-power test, and aligned-peak test consistently require a factor of 2 dilation between the dimmest and brightest BATSE events (Norris 1994, Norris et al 1994) as do the $T_{50}$ and $T_{90}$ distributions (Norris et al. 1995a). Norris et al. (1994, 1995a) interpret the factor of two dilation as consistent with the GRB Log N-Log P distribution, although perhaps with some evolution. Fenimore & Bloom (1995) contends that it is not consistent when one includes all the factors relating distance to time dilation. One key factor involves the tendency for peaks in GRB time histories to be narrower at higher energy. Fishman et al. (1992) noted that individual peaks frequently are narrower and better defined at higher energies. Link, Epstein, and Priedhorsky (1993), showed that this is a prevalent property of most bursts. In this paper we show that there is a well defined relationship for the average width of peaks as a function of energy. We, like Link, Epstein, and Priedhorsky (1993) will use the autocorrelation function of the GRB time history. We will show that the average autocorrelation function for many bursts is a very well behaved function with a shape that is universal. Heuristically, an autocorrelation measures the average relative intensity between points in the time history that are separated by an amount of time called the lag. If a GRB time history is stretched by a factor, the width of an autocorrelation function that covers the entire time history is increased by the same factor, assuming that noise does not dominate. In some cases it is not practical to use the entire time history. We always use enough of the time history such that the actual length that we use does not have a large impact on the analysis (see section 3). As such, it can be used to detect changes in time scales that might be associated with the expansion of the universe or measure the average peak width as a function of energy. The average autocorrelation is fairly immune to systematic effects such as the identification of the highest peak. The noise is explicitly accounted for by calculating the expected autocorrelation given the noise level. The average autocorrelation is similar to the aligned peak tests in that the peak of each burst is used as a fiducial to form an average and is sensitive to time scales the order of a few seconds. However, the autocorrelation uses much more data so it has better statistics. It is roughly equivalent to aligning most of the peaks in a burst rather than just the highest.

## 2. INSTRUMENTATION

The BATSE experiment on CGRO uses eight large area detectors (LADs) to locate and study GRBs over a large dynamic range. Usually, the detector count rate is recorded with modest temporal resolution (1.024 sec). When there is a statistically significant increase in the counting rate above the background, special modes are triggered and the data is recorded at a variety of temporal and energy resolutions. See Fishman et al. (1992) for further details on the instrumentation and data modes. For the purposes of this study, we will use the four channel triggered data. This consists of the counts in four broad energy bins labelled "1" for 25 to 57 keV, "2" for 57 to 115 keV, "3" for 115 to 320 keV, and "4" for above 320 keV which is effectively 320 to 1000 keV. A data set labelled "1+2" combines "1" and "2" together to effectively create a 25 to 115 keV channel. A memory records $\sim 2$ sec of data before the trigger and the duration of the recorded data after the trigger is $\sim 240$ sec. The time resolution for this data is 0.064 sec. In Norris et al. (1994), the period prior to the trigger is augmented by rebinning the continuously available 1.024 sec samples into 0.064 sec samples. This extends the pretrigger by $\sim 16$ sec.

We will use the same data set as used by Norris et al. 1994 including the augmented pre-trigger. The on-board electronics adds data from all the detectors that triggered, this can vary from two to four detectors. Bursts were assigned by Norris et al. (1994) to a brightness class based on the largest net count rate in 0.064 sec samples in channel 1+2+3+4. Those bursts with count rates between 18,000 and 250,000 cts sec$^{-1}$ are called "bright" bursts, those with counts rates between 2,400 and 4,500 cts sec$^{-1}$ are "dim", and those with count rates between 1,400 and 2,400 are called "dimmest". (Events with an intermediate count rate are not used since the time dilation effects are largest for well separated classes.) Short events (defined here and in Norris et al. 1994 to have durations less than 1.5 sec) were excluded from the study. In this paper, we seek the intrinsic variation with energy of the width of the temporal peaks. We use only the bright events to avoid potential effects due to different distances including time dilation from the expansion of the universe. Even if GRBs come from cosmological distances, under the standard candle assumption these events are all from approximately the same distance and therefore have the same stretching due to the expansion of the universe. There were 45 useable events in the bright class. This data set is the same as used by Norris et al. 1994 although processed completely independently with the exception of the selection of events and the augmentation of the pre-trigger data.

## 3. THE AUTO-CORRELATION FUNCTION

The autocorrelation function for GRBs was investigated by Link, Epstein, and Priedhorsky (1993) where it was shown that time scales are almost always shorter at higher energies. Following Link, Epstein, and Priedhorsky, let $m_i$ be the observed gross counts in discretely sampled data in $n$ bins of equal size $\Delta T$ ranging from $-n\Delta T/2$ to $+n\Delta T/2$, about the largest peak in the GRB time history. Here $m_i$ is number of counts, so it follows Poisson statistics. Let $b_i$ be the corresponding background counts. We determined the background by a linear fit to regions before and after the bursts where it was judged by eye to be inactive. The net counts are $c_i = m_i - b_i$ and the estimate of the true autocorrelation as a function of the lag, $\tau = j\Delta T$, is

$$A(\tau) = 1 \qquad\qquad j = 0,$$

$$= \sum_{i=-n/2}^{n/2} \frac{c_{i+j} c_i}{A_0} \qquad -n/2 < j < n/2 \ . \qquad (1)$$

Here, $c_{i+j}$ is zero if $i + j > n/2$ or $i + j < -n/2$. When studying individual events (as Link, Epstein, & Priedhorsky [1993] did), one can always use the entire time history. We average many events together and must use them in a uniform manner. This requires us to select the duration to use. If we selected a very short duration then the auto-correlation is not sensitive to time stretching (a flat-topped burst exceeding the selected duration would show no difference in the auto-correlation). In we selected a very large duration, then there would be many instances when bursts would have to be left out of the averaging because the peak occurs within $\pm\Delta T/2$ such that $c_{i+j}$ is not defined for part of the needed range. Although bursts can be very long, usually emission more than a few seconds away from the largest peak contributes only a little to the auto-correlation function. We have found only small differences for all $n\Delta T > 8$ s and have used $n\Delta t = 16$ s throughout this paper. By definition, the autocorrelation is symmetric, $A(\tau) = A(-\tau)$. The normalization, $A_0$, is

$$A_0 = \sum_{i=-n/2}^{n/2} c_i^2 - m_i \ . \qquad (2)$$

An autocorrelation without normalization would have a large peak at $\tau = 0$ where all the noise adds coherently and would be count rate dependent at $\tau \neq 0$. The $-m_i$ term in equation (2) normalizes the autocorrelation to that expected without noise.

In order to fit functions to the observed autocorrelations, we require a measure of its uncertainty. The terms of $A(\tau)$ are not statistically independent. However, our use of the uncertainty is only to obtain a relative goodness of fit. In fact, each term of the variance on $A(\tau)$ will be approximately the same so its exact value is not important. The variance of the numerator of equation (1) is

$$\sigma^2_{c \star c_j} = \sum_{i=-n/2}^{n/2} c_i^2 |c_{i+j}| + |c_i| c_{i+j}^2 \ . \tag{3}$$

(We assume that the variance propagated from the background is small since the background is based on much more data that the individual points.) The variance on the normalization is

$$\sigma^2_{A_0} = \sum_{i=-n/2}^{n/2} 4c_i^3 + m_i \ . \tag{4}$$

Combining equations (3) and (4) gives the variance on the $j$-th term of the autocorrelation:

$$\sigma^2_{A(j\Delta T)} = \frac{\sigma^2_{c \star c_j}}{A_0^2} + \frac{A^2(j\Delta T)}{A_0^2} \sigma^2_{A_0} \ . \tag{5}$$

The average of a fair number of GRB autocorrelations is quite stable and shows only a small variation. Let $\bar{A}(i, \tau)$ be the average autocorrelation for the $i$-th channel or combination of channels. Figure 1 shows the average of the bright events for the four channels of the LAD data, that is, $\bar{A}(i,\tau) = \sum_{k=1}^{N_B} A_k(i,\tau)/N_B$ where $N_B$ is the number of bright events (45) and $k$ denotes different bursts. The normalization of each autocorrelation is such that each burst contributes equally to the average autocorrelation independent of its brightness. Note how clear the energy dependency is in Figure 1. Figure 1 is semi-log and the curves appear nearly as straight lines so the shape of the autocorrelation is approximately an exponential.

Each energy channel is nearly an exact time stretched version of the others. We define $\mathcal{S}_{i,j}$ to be the best-fit factor that scales $\bar{A}(i,\tau)$ into $\bar{A}(j,\tau)$. It is found by minimizing

$$\chi^2 = \sum_{l=1}^{m} \frac{\left(\bar{A}(j,l\Delta T) - \lambda_{ij}\bar{A}(i,\mathcal{S}_{i,j}l\Delta T)\right)^2}{\sigma^2_{\bar{A}(j,l\Delta T)} + \sigma^2_{\bar{A}(i,l\mathcal{S}_{i,j}\Delta T)}} \ . \tag{6}$$

Here $m\Delta T$ is the range of lags that is used in the fit. This range is set by where the functions are well defined. The auto-correlation of the highest energy channel begins to have significant noise at $\sim 2.5$ s (see Fig. 1) so we have used $m\Delta T = 2.5$ sec. Since $A(\tau)$ is symmetric nothing is gained by including negative lags.

In each panel of Figure 2, we show the average autocorrelation for two channels. For example, in Figure 2a, we show $\bar{A}(1,\tau)$ (i.e., 25 to 57 keV) and $\bar{A}(2,\tau)$ (i.e., 57 to 115 keV). Also shown as a bold curve is the time stretched autocorrelation of the higher energy channel that best fits the lower energy channel (e.g., the bold curve in Fig. 2a is $\lambda_{21}\bar{A}(2,\mathcal{S}_{2,1}\tau)$). The overall scaling ($\lambda$) is always very near unity. Note, for example, in Figure 2c that the bold curve slightly exceeds unity at $\tau = 0$. For energy channels with poorer statistics, the uncertainty of the normalization is reflected in $\sigma^2_{\bar{A}}$ which is nearly constant as a function of $\tau$ since $m\Delta T$ (2.5 sec) is much less than $n\Delta T$ (16 sec). The goodness-of-fit parameter in equation (6) will not follow the $\chi^2$ statistic because the points are not independent. The purpose of $\sigma^2_{\bar{A}}$ is to balance the uncertainty in the overall scale factor ($\lambda$) with the uncertainty in the time stretching ($S$). Figure 2 demonstrates that the average autocorrelation has a universal shape (but different time stretching) for all energies. Note how well the time-stretched higher energy autocorrelation always agrees with the broader (lower energy) autocorrelation. Even the highest energy range (320-1000 keV), which showed a deviation from an exponential in Figure 1 scales exactly into the lower energy autocorrelations.

The curves in Figure 1 are not pure exponentials, there is a slight curve to the histograms. We have tried to fit a variety of functions to the histograms. A single exponential ($\exp^{-\tau/\tau_0}$) fits poorly, especially the higher energy channels. A function such as $\exp^{a\tau^2+b\tau}$ fits the lower energy channels well but not the higher energy channels. Although not unique, the most successful function that we tried is a sum of two exponentials:

$$\bar{A}(i,\tau) = \beta_i \exp^{-|\tau|/\alpha_{i1}} + (1-\beta_i)\exp^{-|\tau|/\alpha_{i2}} \qquad (7)$$

where $i$ denotes the 4 energy channels. To determine the free parameters in equation (7), we minimize

$$\chi^2 = \sum_{j=1}^{n/2} \left[\bar{A}(i,j\Delta T) - \beta_i \exp^{-|\tau|/\alpha_{i1}} - (1-\beta_i)\exp^{-|\tau|/\alpha_{i2}}\right]^2. \qquad (8)$$

The parameters $\alpha_{i1}$ and $\alpha_{i2}$ are found by searching the parameter space and

$\beta_i$ is found analytically from $\delta\chi^2/\delta\beta_i = 0$. The best fits are

$$\bar{A}(1,\tau) = 0.66\exp^{-|\tau|/2.40} + 0.34\exp^{-|\tau|/25}, \qquad (9a)$$

$$\bar{A}(2,\tau) = 0.64\exp^{-|\tau|/1.74} + 0.36\exp^{-|\tau|/25}, \qquad (9b)$$

$$\bar{A}(3,\tau) = 0.48\exp^{-|\tau|/0.94} + 0.52\exp^{-|\tau|/9.9}, \qquad (9c)$$

$$\bar{A}(4,\tau) = 0.53\exp^{-|\tau|/0.56} + 0.47\exp^{-|\tau|/6.5}. \qquad (9d)$$

All values of $\alpha_{i2}$ above 25 are equally consistent with the data. These curves are plotted as curves in Figure 1. Equation (8) does not follow the $\chi^2$ statistic because the $\bar{A}$ terms are not independent so we cannot qualitatively evaluate the fit. However, as seen in Figure 1, the fit is excellent.

## 4. ENERGY DEPENDENCE OF TIME SCALE

We will characterize the energy dependence of the typical time scale in the GRB time history using two different measures: the width of $\bar{A}$ and the width of the average pulse profile (from Norris et al. 1994, 1995b). The solid triangles in Figure 3 are the width ($W_{ac}$) of each autocorrelation from Figure 1 as measured by where $\ln\bar{A}(\tau) = 0.5$. Since Figure 3 is log-log and the points nearly lie on a straight line, we have fit a power law to the points. The best fit power law is

$$W_{ac}(E) = 17.4E^{-0.43}. \qquad (10)$$

This function is shown in Figure 3 as a solid line. This is a robust result. Using the width at other values of $\bar{A}$ gives similar results. Also, the fact that the auto-correlation function for each energy can be scaled into another and they overlap so well (Fig. 2) implies that the power law holds for more than just the point where $\ln \bar{A}(\tau) = 0.5$ (see discussion of eq. [14]).

One thing that is not clear in our formulation is what energy to place the points at. We have placed them at the energy corresponding to the lower energy bound of the channel they represent. The autocorrelation function is quadratic in counts (see eq. [1]) so for any particular channel, the width reflects where most of the counts are. For example, if one combines channel 3 and 4, it has effectively the same width as 3 only. If we were to use the midpoint of the channel, it is still a power law: $W_{ac} = 18.1 E^{-0.40}$. Another possible measure of the time scale of $\bar{A}$ is how much one energy range needs to be stretched to map it into another energy range. This is not independent from $W_{ac}$ but serves as another way to measure it (eq. [6] rather than eq. [1]). Using channel 1 as a baseline (i.e., if $S_{11} = 1$), $S_{21}^{-1} = 0.78$, $S_{31}^{-1} = 0.54$, and $S_{41}^{-1} = 0.33$ give how much the autocorrelations of the higher energy channels are narrower as a function of energy. Fitting the $S_{i1}^{-1}$ points gives

$$S_{i1}^{-1} = 4.45 E_i^{-0.46}. \qquad (11)$$

A second measure of the time scale of GRBs comes from the average pulse width. Norris et al. 1994 decomposed GRB time histories into individual pulses and found the average rise and fall time scales. The width (in seconds) of the rise/fall of the average pulse profile are 0.22/0.44, 0.17/0.32, 0.13/0.27, and 0.08/0.18 for BATSE channels 1, 2, 3, and 4, respectively. We plot in Figure 3 as squares the sum of the rise and fall times as a function of energy. Again, the energy dependence of the time scale appears to be a power law.

In this case, the average pulse width is

$$W_{ap} = 2.1 E^{-0.37}, \qquad (12)$$

which is plotted in Figure 3 as a dashed line. Another measure of the pulse width is the average full width, half maximum (FWHM). Norris et al. (1995b) reports that the average FWHM for the four BATSE channels are 0.817, 0.616, 0.473, and 0.287 sec. These widths can be fit by a power law as well:

$$W_{FWHM} = 3.2 E^{-0.42}. \qquad (13)$$

This, too, is very robust. Norris et al. (1995b) reports the width for seven different fractions of the peak height and all seven can be fit with a power law.

We note that the average pulse width is much less than the width from the autocorrelation function. The individual pulses are narrower than the clusters of peaks that often determine the autocorrelation width. However, there is not a simple relationship between the two measures. For example, simulations of shot noise with pulses the order of the average pulse width produce autocorrelation functions that are much narrower than observed.

In equation (8), we fit each energy channel separately using 12 parameters. From Figure 3, we see that the energy dependence of the time scale in GRBs is a power law. The parameters $\alpha_{i1}$ and $\alpha_{i2}$ found in equation (9) do not follow a power law. However, it is possible to have a functional form that has a power law dependency on energy and fits within the noise. We fit all four curves in Figure 1 with

$$\bar{A}(i,\tau) = \beta \exp^{-\frac{|\tau|}{k_1 E_i^{-\alpha}}} + (1-\beta) \exp^{-\frac{|\tau|}{k_2 E_i^{-\alpha}}}. \qquad (14)$$

This form accommodates the three characteristics of the average autocorrelation: it consists of two exponentials, the energy dependence scales as a power law, and the shape of one energy range scales linearly with time into all the other energy ranges. In Figure 4 we show the autocorrelations from Figure 1 fit with equation (14). The best fit parameters are $\beta = 0.55$, $k_1 = 8.75$, $k_2 = 154$, and $\alpha = 0.45$. Although the curves deviate some from the histograms, the fits are within the uncertainty expected for the average autocorrelation caused by variations in the choice of bursts that are included in the average. Thus, we consider the fit acceptable.

In summary, we find that the average autocorrelation of GRBs time histories is a universal function that can measure the time scale as a function of energy. The dependence is a power law in energy with an index that is between 0.37 and 0.46, depending on how it is measured. This is the first quantitative relationship between temporal and spectral structure in gamma-ray bursts. The energy dependence is important for two reasons. First, the shape may indicate the underlying physics responsible for the time history. For example, the subpeak's temporal width might be produced by the growth of a shock within a relativistic expanding shell in a cosmological GRB and the power law dependence on energy is related to how the shock converts bulk motion into gamma-rays. Alternatively, the energy dependence might be related to how a disturbance propagates on the surface of a neutron star. Second, in order to interpret the time dilation due to the expansion of the universe, one must understand the energy dependence which competes with the cosmological time dilation to form the temporal width.

Fenimore & Bloom (1995) includes the energy dependence in the interpretation of time stretching as a function of burst brightness and concludes that the observed time dilation is not consistent with the observed log $N$-log $P$ distribution (Fenimore et al. 1993) unless there is strong evolution and it is only coincidental that the log $N$-log $P$ distribution shows a -3/2 power law..

*Acknowledgement* This work was done under the auspices of the US Department of Energy and was funded in part by the Compton Gamma-Ray Observatory Guest Investigator program.

FIGURE CAPTIONS

**Fig. 1** The average autocorrelation of 45 bright BATSE gamma-ray bursts in four energy channels. At higher energies, gamma-ray bursts have shorter time scales. The solid curves are fits of the sum of two exponentials to the autocorrelation histograms.

**Fig. 2:** Comparisons of pairs of energy channels for the BATSE bright events. Each panel shows the average autocorrelation for two energy channels from the BATSE LAD data. The bold histogram is a best fit, time stretched version of the narrower (higher energy) channel fit to the broader (lower energy) channel. The average autocorrelation apparently has a universal shape which is approximately exponential.

**Fig. 3:** The energy dependency of the time scale as determined by the average autocorrelation and the average pulse width. The triangles are the half width of the autocorrelation function and the solid curve is a best fit to the triangles. The squares are the sum of the rise and fall of the average pulse profile and the dashed line is a best fit to it. In each case, the time scale of the temporal structure within the GRB scales as a power law of the energy with an index of $\sim$ -0.4.

**Fig. 4** The average autocorrelation of 45 bright BATSE gamma-ray bursts in four energy channels fit with the sum of two exponentials where widths scale as a power law.

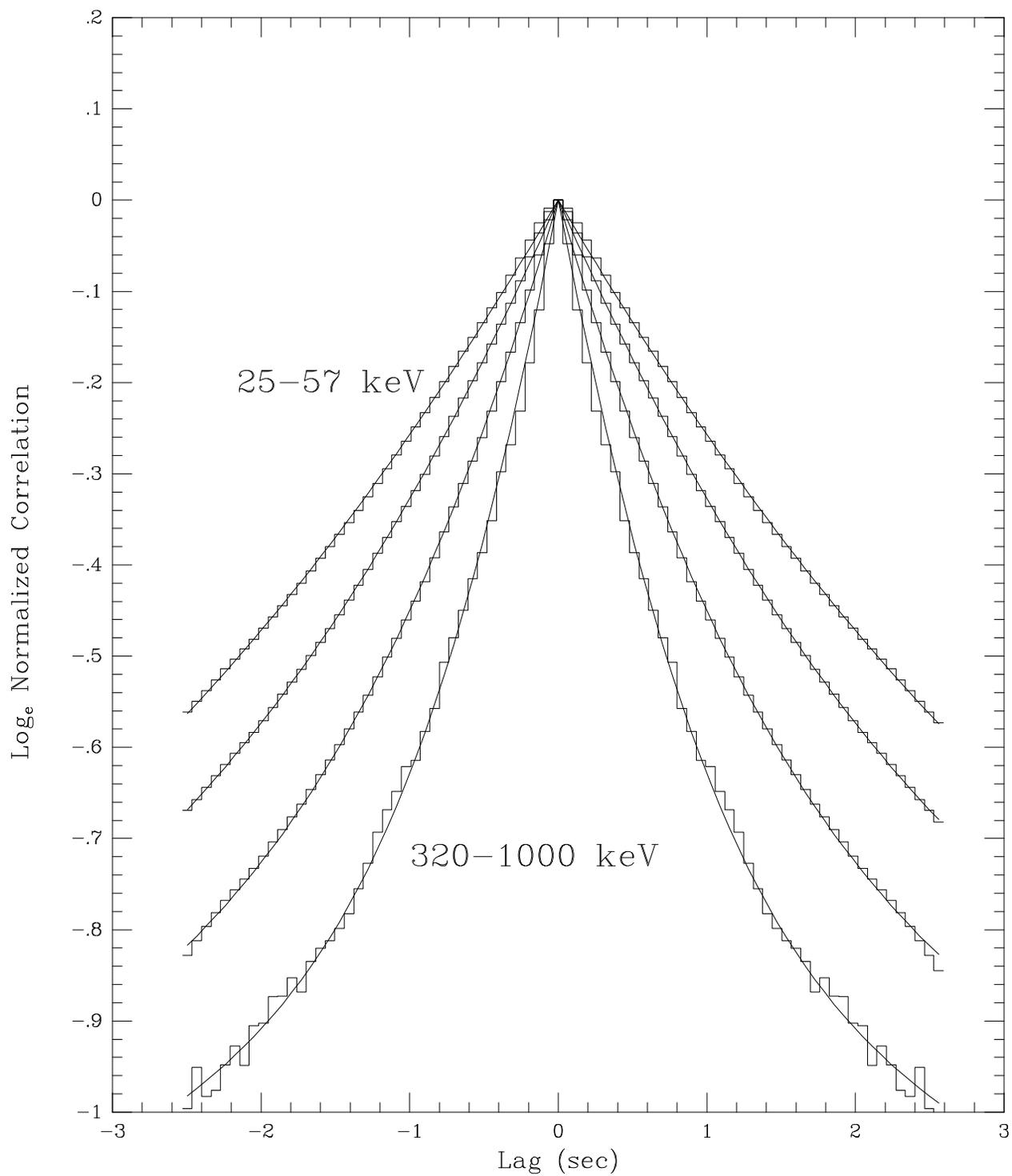

Fig. 1: Energy Dependency of BATSE Bright

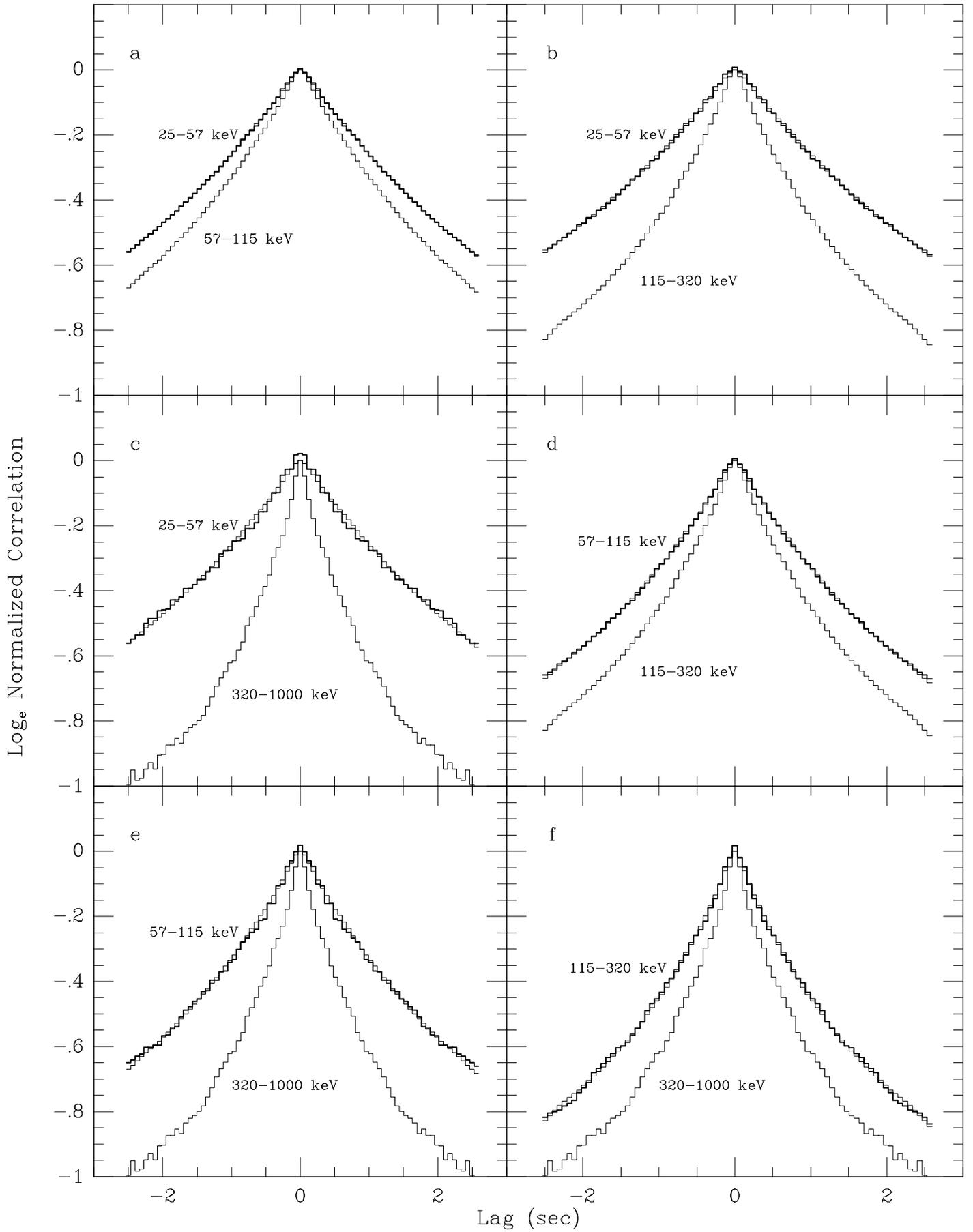

Fig 2

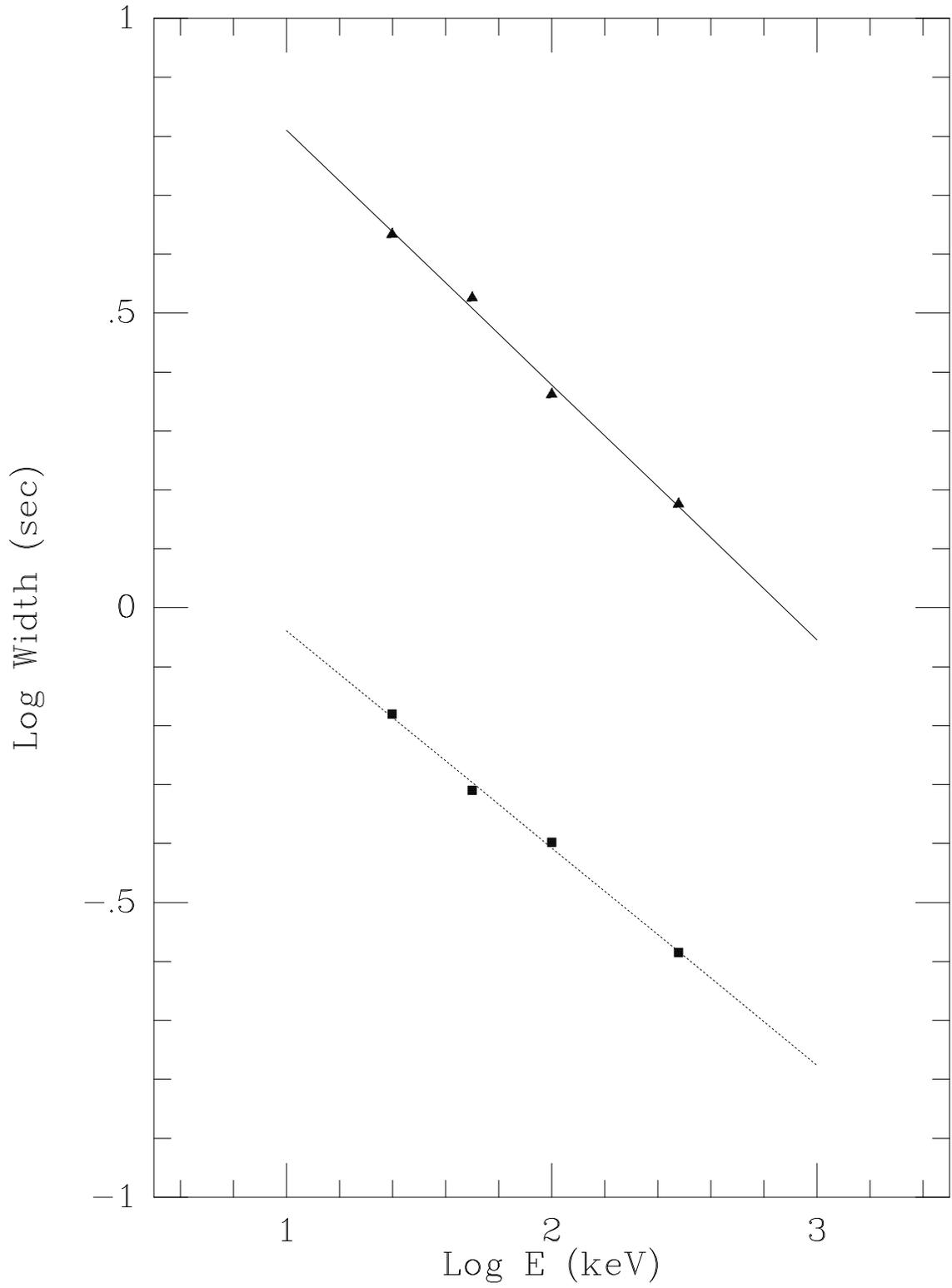

Figure 3

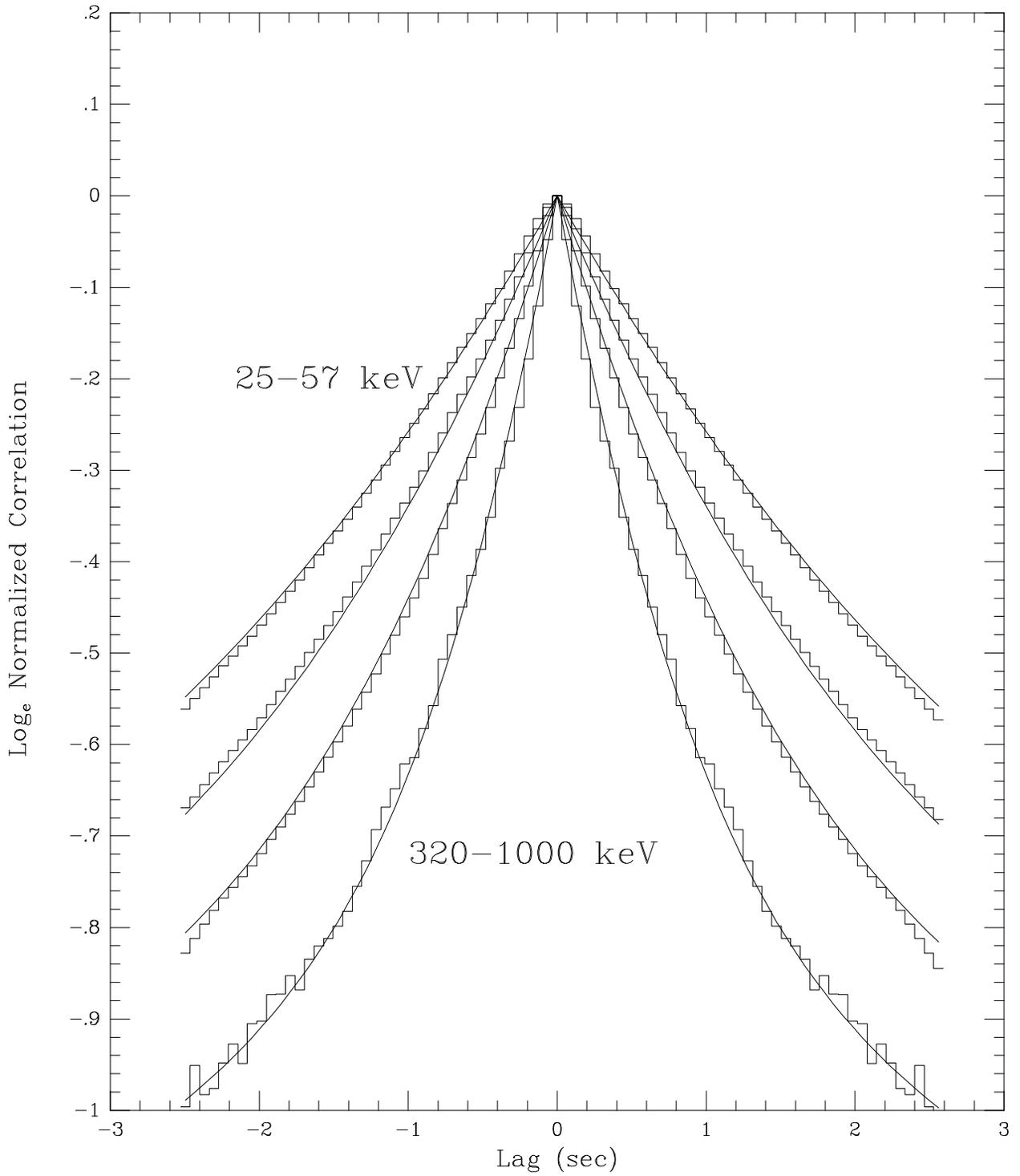

Fig. 4: Energy Dependency of BATSE Bright

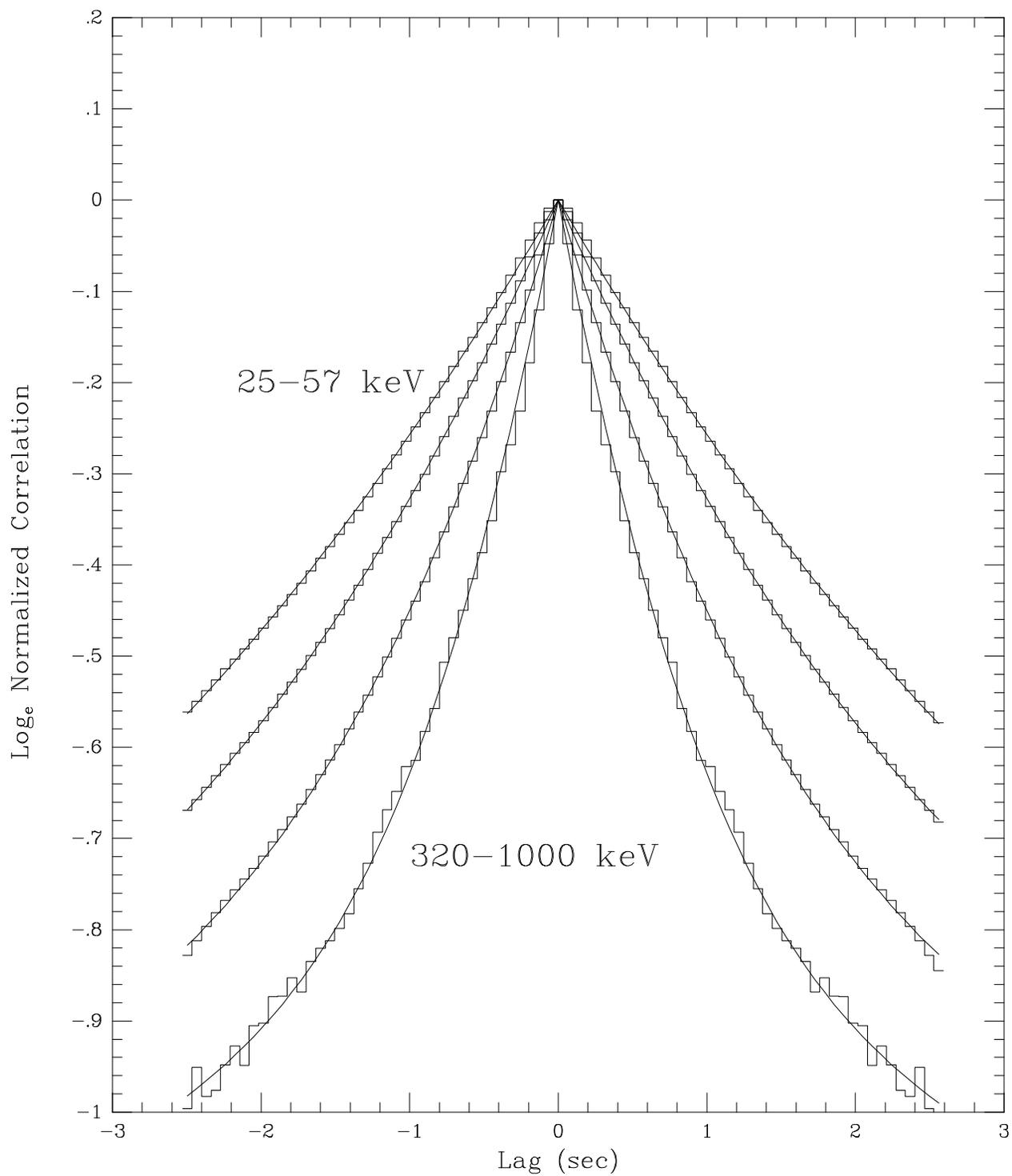

Fig. 1: Energy Dependency of BATSE Bright

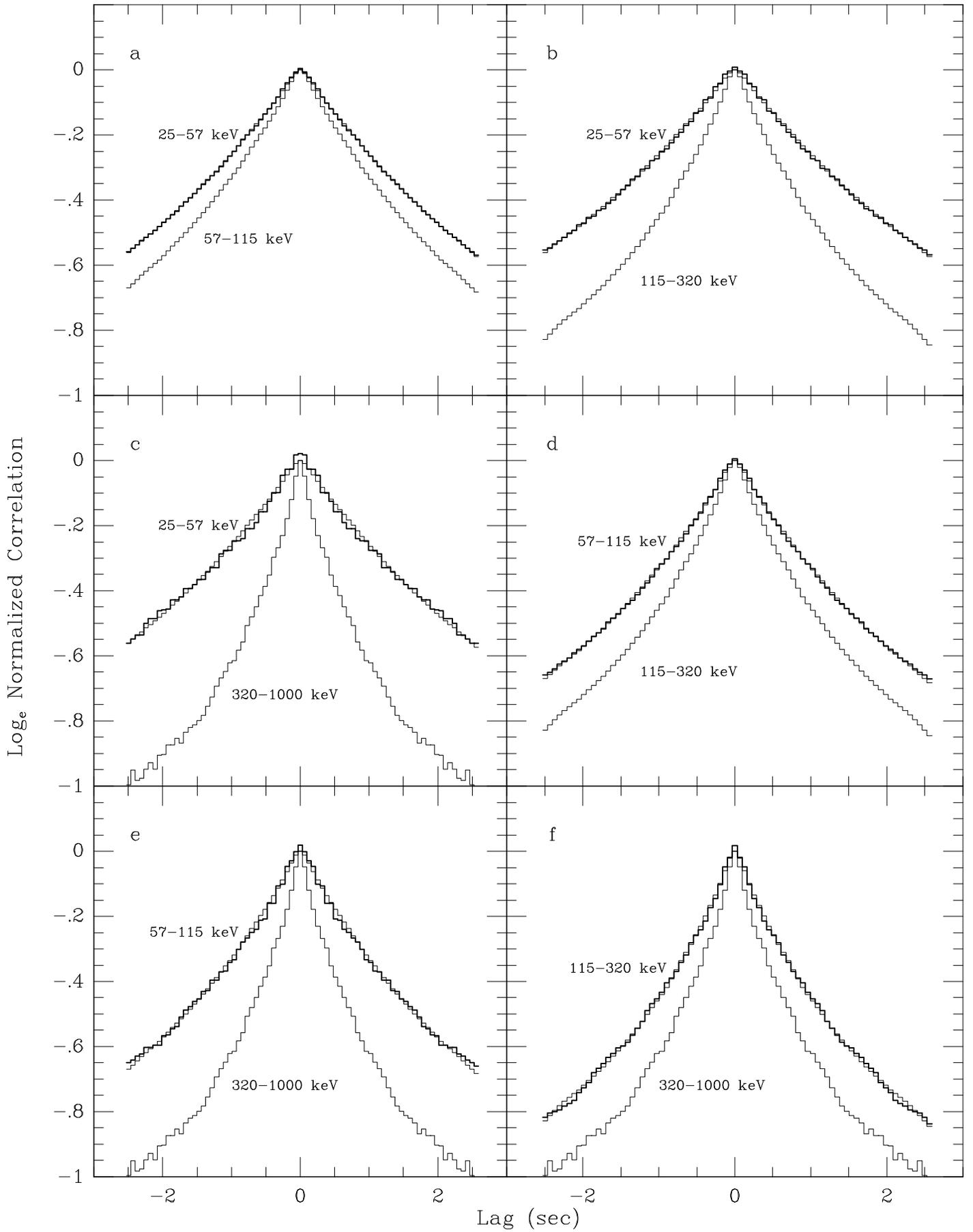

Fig 2

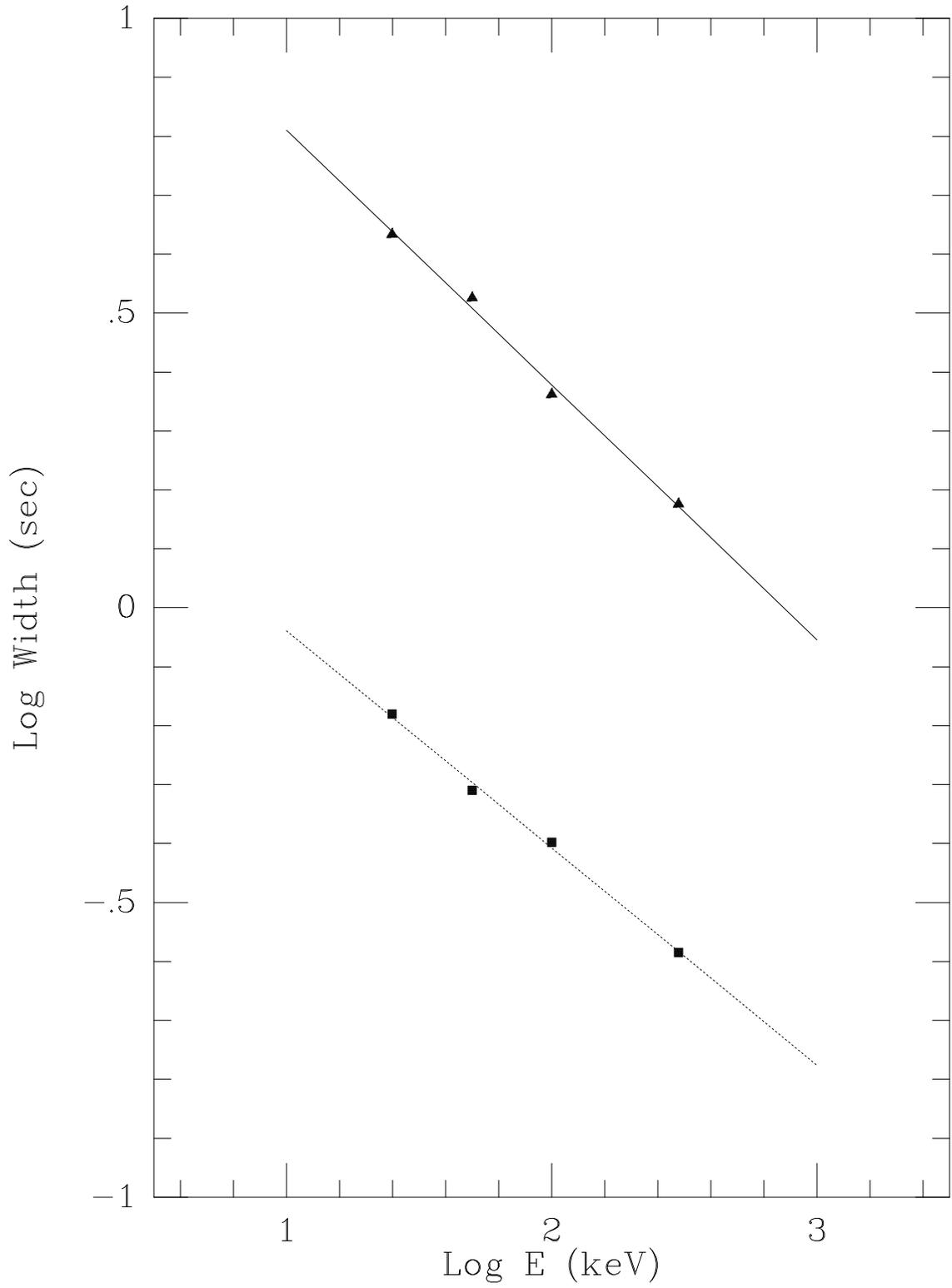

Figure 3

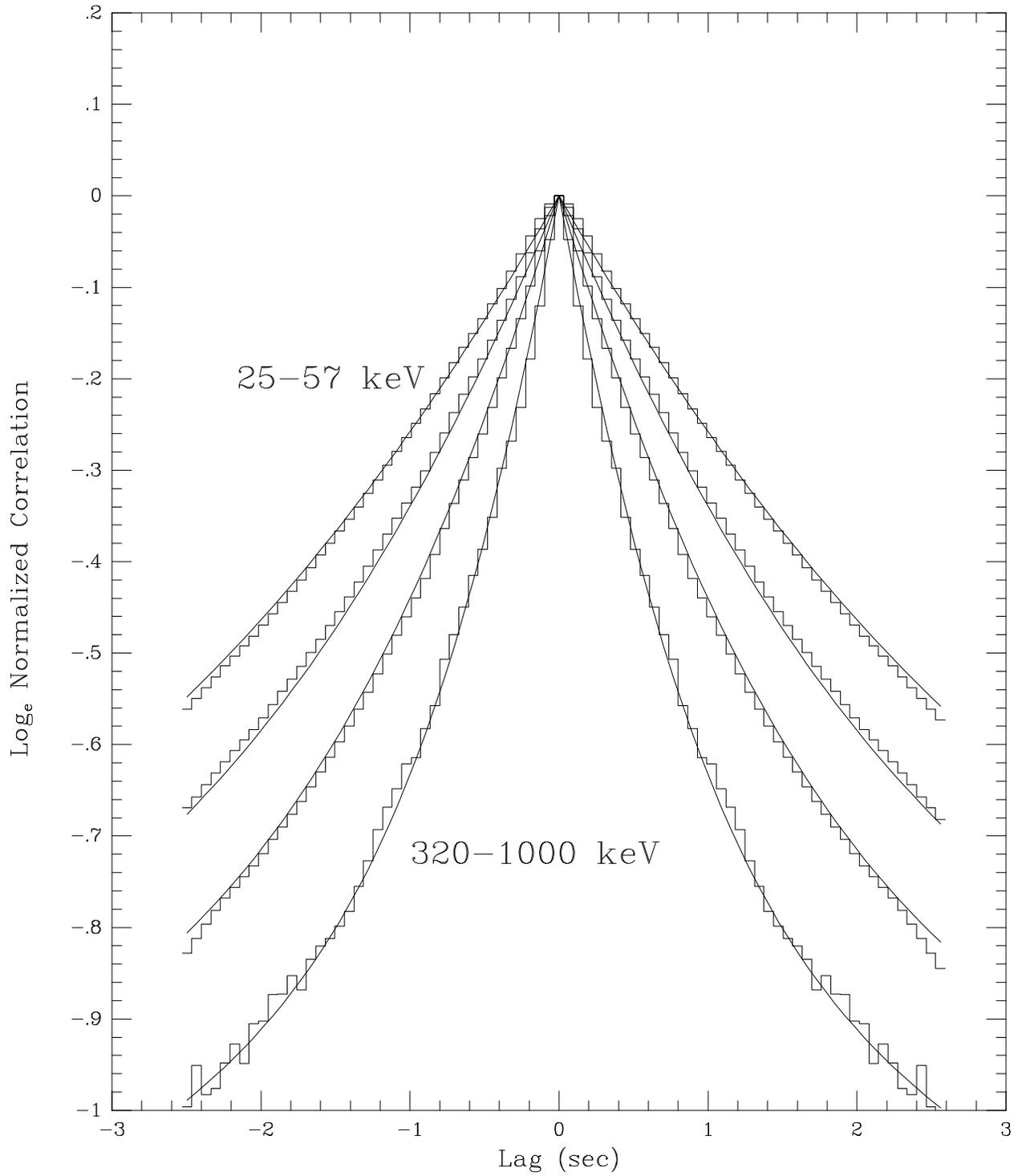

Fig. 4: Energy Dependency of BATSE Bright